\documentclass[12pt]{article}
\begin{document}
\title{The Hamilton-Jacobi treatment for non-abelian Chern-Simons
system}\maketitle
\begin{center}
\author{S. I. Muslih\\{Department of Physics,
Al-Azhar University, Gaza, Palestine}}
\end{center}
\begin{abstract}
The non-abelian Chern-Simons field interacting with $N$ component
complex field is treated as a constrained system using the
Hamilton-Jacobi approach. The reduced phase space Hamiltonian
density is obtained with out introducing Lagrange multipliers and
with out any additional gauge fixing condition. The quantization
of this system is also discussed.
\end{abstract}
\newpage
\setcounter{section}{0}

\section{Introduction}

 In an earlier work \cite{sm1}, it was shown that the canonical method \cite{sm2,gl} leads  to obtain the equations of
motion and the correct reduced phase space Hamiltonian for an
abelian Chern-Simons field system interacting with scalar field
without using any gauge fixing conditions.

In the present paper we summarize our results for a similar
demonstration for the non-abelian Chern-Simons term. In fact,
this system has been discussed by several authors
\cite{fer,des,shar} and it was shown in reference \cite{shar},
that the gauge conditions consistent with equations of motion can
be chosen. We show in this work that using the Hamiton-Jacobi
method \cite{sm2,gl}, the reduced phase space Hamiltonian density
is obtained without using any gauge fixing conditions. Besides, we
discuss the quantization of this system using the Hamilton-Jacobi
method.

\section{The Hamilton-Jacobi method of constrained systems}
In this section we will briefly review the Hamilton-Jacobi
method  \cite{sm2,gl} for studying the constrained systems.

This formulation leads us to obtain the set of Hamilton-Jacobi
partial differential equations [HJPDE] as follows:
\begin{eqnarray} &&H^{'}_{\alpha}\left(t_{\beta}, q_a, \frac{\partial
S}{\partial q_a},\frac{\partial S}{\partial t_{\alpha}}\right)
=0,\nonumber\\&&\alpha,\;\; \beta=0,\;n-r+1,...,n,
\nonumber\\&&a=1,...,n-r,
\end{eqnarray}
 where
\begin{equation}
H^{'}_{\alpha}=H_{\alpha}(t_{\beta}, q_a, p_a) + p_{\alpha},
\end{equation}
and the Hamiltonian $H_{0}$ is defined as
\begin{eqnarray}
 &&H_{0}= p_{a}w_{a}+
p_{\mu}
 \dot{q_{\mu}}|_{p_{\nu}=-H_{\nu}}\nonumber\\&&-
L(t, q_i, \dot{q_{\nu}},
\dot{q_{a}}=w_a),\nonumber\\&&\;\;\;\;\;\;\;\;\mu,~\nu=n-r+1,...,n.
\end{eqnarray}

The equations of motion are obtained as total differential
equations in many variables as follows \cite{gl}:
\begin{eqnarray}
 &&dq_a=\frac{\partial H^{'}_{\alpha}}{\partial
 p_a}dt_{\alpha},\;\;\;\;
 dp_a= -\frac{\partial H^{'}_{\alpha}}{\partial q_a}dt_{\alpha},\\
&&dp_{\beta}= -\frac{\partial H^{'}_{\alpha}}{\partial t_{\beta}}dt_{\alpha},\\
&& dz=\left(-H_{\alpha}+ p_a \frac{\partial
H^{'}_{\alpha}}{\partial p_a}\right)dt_{\alpha},
\end{eqnarray}
where $z=S(t_{\alpha};q_a)$. The set of equations (4-6) is
integrable
 if \cite{sm2}
\begin{equation}
dH^{'}_{0}=0,\;\;\;\;\;dH^{'}_{\mu}=0.
\end{equation}
If condition (7) are not satisfied identically, one considers
them as new constraints and again testes the consistency
conditions. Hence, the canonical formulation leads to obtain the
set of canonical phase space coordinates $q_a$ and $p_a$ as
functions of $t_{\alpha}$, besides the canonical action integral
is obtained in terms of the canonical coordinates. The
Hamiltonians $H^{'}_{\alpha}$ are considered as the infinitesimal
generators of canonical transformations given by parameters
$t_{\alpha}$ respectively.
\section{ Non-abelian Chern-Simons as a constrained system}
 The Lagrangian density for Chern-Simons field
interacting a scalar field in $2+1$ dimensions is given by
\begin{eqnarray}
&&{\cal L}=
[D_{\mu}{\Phi}_{\alpha}]^{\dag}[{D^{\mu}}\Phi_{\alpha}]\nonumber\\&&+
{\frac{\kappa}{4\pi}}{\epsilon}^{{\mu}{\nu}{\lambda}}
Tr\left[A_{\mu}{\partial_{\nu}}A_{\lambda}
-\frac{2}{3}A_{\mu}A_{\nu}A_{\lambda}\right],
\end{eqnarray}
where $\Phi$ is $N$- component scalar field which transforms
according to fundamental relation of Lie group $SU(N)$ and the
parameter $\kappa$ plays the role of a coupling constant. Here
\begin{eqnarray} &&D_{\mu}= \partial_{\mu} -i eA_{\mu},\\
&&A_{\mu} = i gT_{a} A^{a}_{\mu},
\end{eqnarray}
and $T^{a}$ are the generators of the compact Lie group $SU(N)$
\begin{equation}
[T^{a}, T^{b}]_{-}= i f^{abc}T^{c},
\end{equation}
\begin{equation}
[T^{a}, T^{b}]_{+} = d^{abc}T^{c} + \frac{1}{N}\delta^{ab},
\end{equation}
\begin{equation}
Tr(T^{a}T^{b})=\frac{1}{2}\delta^{ab},
\end{equation}
where $d^{abc}$ is totally symmetric tensor in the indices $a, b,
c$ while $f^{abc}$ are the usual structure constants of $SU(N)$.

Canonically conjugated momenta for $\Phi$ and $\Phi^{\dag}$ are
called $P$ and $P^{\dag}$, where
\begin{equation}
P_{\alpha} = {\dot \Phi}_{\alpha}^{\dag} + ig
{\Phi}_{\beta}^{\dag}A_{0}^{a}T_{\beta\alpha}^{a}.
\end{equation}
The canonical momenta $\Pi_{\rho}^{a}$ conjugated to
$A_{\rho}^{a}$ are defined as
\begin{equation}
\Pi_{\rho}^{a}= -\frac{\kappa}{8\pi}g^{2} {\epsilon}^{\mu 0
\rho}A_{\mu}^{a}.
\end{equation}

 The primary constraints are obtained as follows:
\begin{eqnarray}
&&{\cal H'}_{0}^{a}= \Pi_{0}^{a} =0,\\
&&{\cal H'}_{1}^{a}=  \Pi_{1}^{a} +
\frac{\kappa}{8\pi}g^{2}A_{2}^{a}=0,\\
&&{\cal H'}_{2}^{a}= \Pi_{2}^{a} -
\frac{\kappa}{8\pi}g^{2}A_{1}^{a}=0.
\end{eqnarray}

The canonical Hamiltonian density can be written as
\begin{eqnarray}
&&{\cal H}_{0}^{c} = P_{\alpha}^{\dag}P_{\alpha}+
(\partial_{i}\Phi_{\alpha}^{\dag})(\partial^{i}\Phi_{\alpha})\nonumber\\&&+
A_{0}^{a}j_{0}^{a} +A_{i}^{a}j_{i}^{a} +
g^{2}\Phi_{\beta}^{\dag}A_{i}^{a}T_{\beta\alpha}^{a}T_{\alpha\gamma}^{b}A_{i}^{b}\Phi_{\gamma}\nonumber\\&&
+\frac{\kappa}{8\pi}g^{2}\epsilon^{ij}[A_{0}^{a}\partial_{i}A_{j}^{a}
+ A_{i}^{a}\partial_{j}A_{0}^{a}]\nonumber\\&& -
\frac{\kappa}{8\pi}g^{3}\epsilon^{ij}
A_{0}^{a}A_{i}^{c}A_{j}^{b}f^{abc},
\end{eqnarray}
where
\begin{eqnarray}
&&j_{0}^{a}= i g[P_{\alpha} \Phi_{\beta}-
P_{\alpha}^{\dag}\Phi_{\beta}^{\dag}]T_{\beta\alpha}^{a},\\
&&j_{i}^{a}= -i g[(\partial_{i} \Phi_{\alpha}^{\dag})\Phi_{\beta}
-
\Phi_{\beta}^{\dag}(\partial_{i}\Phi_{\alpha})]T_{\beta\alpha}^{a}.
\end{eqnarray}

Now, the canonical method leads us to obtain the set of
Hamilton-Jacobi partial differential equations as follows:
\begin{eqnarray}
&&{H'}_{0}^{c}= \int d^{2}x(p_{0} + P_{\alpha}^{\dag}P_{\alpha}+
(\partial_{i}\Phi_{\alpha}^{\dag})(\partial^{i}\Phi_{\alpha})\nonumber\\&&+
A_{0}^{a}j_{0}^{a} +A_{i}^{a}j_{i}^{a} +
g^{2}\Phi_{\beta}^{\dag}A_{i}^{a}T_{\beta\alpha}^{a}T_{\alpha\gamma}^{b}A_{i}^{b}\Phi_{\gamma}\nonumber\\&&
+\frac{\kappa}{8\pi}g^{2}\epsilon^{ij}[A_{0}^{a}\partial_{i}A_{j}^{a}
+ A_{i}^{a}\partial_{j}A_{0}^{a}]\nonumber\\&& -
\frac{\kappa}{8\pi}g^{3}\epsilon^{ij}
A_{0}^{a}A_{i}^{c}A_{j}^{b}f^{abc} =0,\\
&&{H'}_{0}^{a}= \int d^{2}x(\Pi_{0}^{a}) =0,\\
&&{H'}_{1}^{a}= \int d^{2}x\left(\Pi_{1}^{a} +
\frac{\kappa}{8\pi}g^{2}A_{2}^{a}\right)=0,\\
&&{H'}_{2}^{a}=\int d^{2}x\left( \Pi_{2}^{a} -
\frac{\kappa}{8\pi}g^{2}A_{1}^{a}\right)=0.
\end{eqnarray}

To check whether this model is integrable or not we have to
consider the integrabilty conditions (7). In fact, the total
variations of the constraints (22-25) lead to obtain the total
variations of the fields $A_{1}^{a}$ and $A_{2}^{a}$ in terms of
$dt$. Besides, the total variations of ${H'}_{0}^{a}$ yield
\begin{equation}
d{H'}_{0}^{a}={H'}_{3}^{a}dt,
\end{equation}
where ${H'}_{3}^{a}$ are given by
\begin{equation}
{H'}_{3}^{a}=j_{0}^{a} +
\frac{\kappa}{4\pi}g^{2}\epsilon^{ij}[\partial_{i}A_{j}^{b}
-\frac{g}{2}A_{i}^{c}A_{j}^{e}f^{abc}].
\end{equation}
The total variations of ${H'}_{3}^{a}$ vanish and no further
constraints arise.

Now the reduced phase space Hamiltonian density is calculated as
\begin{eqnarray}
&&{\cal H}_{0}^{{c}_{r}}= {\cal H}_{0}^{c}|_ {{H'}_{3}^{a}}=
P_{\alpha}^{\dag}P_{\alpha}+
(\partial_{i}\Phi_{\alpha}^{\dag})(\partial^{i}\Phi_{\alpha})\nonumber\\&&\;\;\;\;\;+
A_{i}^{a}j_{i}^{a} +
g^{2}\Phi_{\beta}^{\dag}A_{i}^{a}T_{\beta\alpha}^{a}T_{\alpha\gamma}^{b}A_{i}^{b}\Phi_{\gamma}.
\end{eqnarray}

 Finally, to obtain the path integral quantization \cite{sm3} of this
model, we have to discuss the integrability conditions in terms
of the action. In fact, Eq. (6) yields
\begin{equation}
dS= \int d^{2}x\left(-{\cal H}_{\alpha}+ p_a \frac{\delta {\cal
H}^{'}_{\alpha}}{\delta p_a}\right)dt_{\alpha}.
\end{equation}
Making use of (29) and (22-25), we obtain the action of the form
\begin{eqnarray}
&&S=\int d^{2}x (- {\cal H}_{0}^{c} dt + \Pi_{0}^{a} dA_{0}^{a} +
\Pi_{1}^{a}dA_{1}^{a}\nonumber\\&& \;\;\;\;\;\;\;\;\;\; +
\Pi_{2}^{a}dA_{2}^{a}+ P_{\alpha}d\Phi_{\alpha} +
P_{\alpha}^{\dag} d\Phi_{\alpha}^{\dag}).
\end{eqnarray}
Using the definition of the canonical momenta, one can recover
the original action
\begin{equation} S=\int d^{3}x {\cal L},
\end{equation}
where ${\cal L}$ is exactly the same as the first-order
Lagrangian (1) in the symplectic formalism.

\section{ Conclusion}
The basic idea of Dirac's method \cite{di} to investigate the
non-abelian Chern-Simons system is to consider the total
Hamiltonian composed by adding the constraints multiplied by
Lagrange multipliers to the canonical Hamiltonian. In order to
derive the equations of motion, one needs to redefine these
unknown multipliers in an arbitrary way \cite{fer,des,shar}.
However, in the Hamilton-Jacobi method \cite{sm2,gl}, there is no
need to introduce Lagrange multipliers to the canonical
Hamiltoian as well as no need to use any gauge fixing conditions.

Unlike conventional methods one can  perform the path integral
quantization of this system using the canonical path integral
method \cite{sm3}to obtain the action directly without
 considering any Lagrange multipliers and without any gauge fixing conditions.

\end{document}